\documentclass[twocolumn,english,aps,prb,reprint,superscriptaddress]{revtex4-2}
\usepackage{amsmath}
\usepackage[utf8]{inputenc}
\usepackage{txfonts}
\usepackage{amssymb}
\usepackage{graphicx}
\usepackage{xcolor}
\usepackage{mathrsfs}
\usepackage{braket}
\usepackage[colorlinks=true, urlcolor=blue, linkcolor=blue, citecolor=blue]{hyperref}
\usepackage{babel}
\usepackage{comment}
\usepackage[T1]{fontenc}

\global\long\def\bk{\mathbf{k}}%

\global\long\def\br{\mathbf{r}}%

\global\long\def\bR{\mathbf{R}}%

\global\long\def\la{\langle}%

\global\long\def\ra{\rangle}%

\makeatletter

\usepackage{babel}

\def\eV{{\,\mathrm{eV}}}
\def\meV{{\,\mathrm{meV}}}

\definecolor{lb}{rgb}{0.000, 0.400, 1.000}

\makeatother

\begin{document}
\title{Symmetry and Minimal Hamiltonian of Nonsymmorphic Collinear Antiferromagnet MnTe}

\author{Koichiro Takahashi}
\thanks{These authors contributed equally to this work.}
\affiliation{Department of Physics and Astronomy, University of New Hampshire,
Durham, New Hampshire 03824, USA}
\author{Hong-Fei Huang}
\thanks{These authors contributed equally to this work.}
\affiliation{School of Physical Science and Technology, 
Soochow University, Suzhou 215006, China}
\author{Jie-Xiang Yu}
\email{jxyu@suda.edu.cn}
\affiliation{School of Physical Science and Technology, 
Soochow University, Suzhou 215006, China}
\author{Jiadong Zang}
\email{jiadong.zang@unh.edu}
\affiliation{Department of Physics and Astronomy, University of New Hampshire,
Durham, New Hampshire 03824, USA}

\begin{abstract}
$\alpha$-MnTe, an $A$-type collinear antiferromagnet, has recently attracted significant attention due to its pronounced spin splitting despite having net zero magnetization, a phenomenon unique for a new class of magnetism dubbed altermagnetism. In this work, we develop a minimal effective Hamiltonian for $\alpha$-MnTe based on realistic orbitals near the Fermi level at both the $\Gamma$ and $A$ points. Our model is derived using group representation theory, first-principles calculations, and tight-binding modeling. The resulting effective Hamiltonian exhibits qualitatively distinct electron transport characteristics between these high-symmetry points and for different in-plane N\'{e}el vector orientations along the $[11\bar{2}0]$ and $[1\bar{1}00]$ directions. Although relativistic correction of the spin-orbit coupling (SOC) is believed to be not important in altermagnets, we show the dominant role of SOC in the spin splitting and valence electrons of MnTe. These findings provide critical insights into altermagnetic electron transport in MnTe and establish a model playground for future theoretical and experimental studies.
\end{abstract}

\maketitle

\section{Introduction}
Manipulation and detection of magnetic order parameters in antiferromagnets have been a challenge in spintronics for decades~\cite{Baltz_2018,Smejkal_2022a}.
The main difficulty is that most collinear antiferromagnets respect a combination symmetry operation of 
a half-translation transformation $t_{1/2}$ or an inversion operation $\mathcal{P}$ with a time reversal transformation $\mathcal{T}$, called 
 the $t_{1/2}\mathcal{T}$ or the $\mathcal{PT}$ symmetry, 
leading to trivial transport properties, 
such as the absence of intrinsic anomalous Hall effect and spin transfer torque.
Besides conventional antiferromagnets, 
altermagnets, as a new variant of antiferromagnets, 
possess collinear N\'eel order parameters with zero net magnetization but break the $t_{1/2}\mathcal{T}$ and $\mathcal{PT}$ symmetry~\cite{YinGen_Theory_MnTe, Smejkal_2022, Smejkal_2022b}.
The non-relativistic spin group of an altermagnet is distinct from that of both ferromagnetism and conventional antiferromagnetism~\cite{Yuan_2020, Yuan_AFM_Classification, XiaobingChen_SpinSpaceGroup, ZhenyuXiao_SpinSpaceGroup},
so that it leads to the bands of the opposite-spin sublattices in reciprocal space not coinciding and can only be connected through a real-space rotation transformation.
Thus, altermagnets are fundamentally distinct from conventional collinear antiferromagnets.
The combination of the zero net magnetization of antiferromagnets with the spin-splitting characteristics of ferromagnets
enables altermagnets not only to generate $\mathcal{T}$-odd spin transport characteristics, 
including intrinsic anomalous Hall effect~\cite{Smejkal_2022a, Feng_2022, Gonzalez_2023, Wang_2023a, Han_2024, Reichlova_2024, Smejkal_2020, Kluczyk_2024},
spin-splitting torque~\cite{GonzalezHernandez_2021, Bose_2022, Bai_2022, Karube_2022},
and tunneling magnetoresistance~\cite{Smejkal_2022c}, 
but also to achieve magnetic stability due to the absence of demagnetizing fields and ultrafast spin dynamic features at terahertz scales~\cite{Zhang_2024, Zhang_2024a, Leenders_2024}.

Spin-orbit coupling (SOC) is often considered to be irrelevant or insignificant for the electronic and magnetic properties in an altermagnet
because the non-relativistic spin splitting with zero net magnetization is regarded as the main characteristics therein~\cite{Smejkal_2020,GonzalezHernandez_2021, Smejkal_2022}.
The type of spin splitting in altermagnets, such as planer or bulk $d$-, $g$-, and $i$-wave, should be uniquely determined by the type of collinear antiferromagnetic ordering and the crystal symmetry in an altermagnet~\cite{Roig_2024}.
However, SOC does play a crucial role in many antiferromagnets not only for the topological non-trivial band structures~\cite{Mong_2010, Wang_2017, Smejkal_2018, Chang_2023}  
but also for the detection and manipulation of the N\'{e}el-vector orientation~\cite{Wadley_2016,Barthem_2013,Bodnar_2018,Zhou_2025, YinGen_Theory_MnTe}.
SOC is known to induce spin mixing and band splitting, such as Rashba-Dresselhaus effect~\cite{Rashba_2015, Dresselhaus_1955} in non-magnetic materials with the inversion symmetry breaking.
The locking between electron spin and momentum under SOC leads to spin-galvanic effect with the uniform spin accumulation.
In an antiferromagnet, this effect can generate opposite spin-orbital torques on the opposite-spin sublattices, achieving the controlling of antiferromagnetism~\cite{Wadley_2016,Bodnar_2018}.
Furthermore, magnetic anisotropy brought about by SOC can alter the symmetry originally determined by the non-relativistic antiferromagnetic system.
To this end, analyzing the non-relativistic magnetic ordering and crystal symmetry is not sufficient to catch the full features of the spin splitting in the band structure.
Both the strength of SOC and magnetic anisotropy is important in antiferromagnetic and altermagnetic spintronics.

$\alpha$-MnTe possesses both the altermagnetic characteristics with the absence of $t_{1/2}\mathcal{T}$/$\mathcal{PT}$ symmetry and the strong SOC originating from the Te atoms, 
so that it has recently garnered widespread attention.
Early studies~\cite{Adachi_theory_NiAsType,Squire_Exp_MnTe, Uchida_Exp_MnTe, Komatsubara_Exp_MnTe, EfremDSa_Exp_MnTe} indicated that bulk MnTe is a $p$-type semiconductor and layered in-plane antiferromagnetic ordering with the N\'{e}el temperature $T_N\approx310$K.
Recently, numerous interesting electronic, magnetic and transport phenomena have been found in MnTe,
including large spin splitting~\cite{FariaJunior_MnTe_EffHamA, Gonzalez_2023, Osumi_2024, Hajlaoui_2024, Belashchenko_2025},
lifting of Kramers spin degeneracy~\cite{YinGen_Theory_MnTe, Krempaský_ARPES_MnTe, Lee_2024, Hajlaoui_2024}, 
the controllable magnetic anisotropy~\cite{Moseley_2022, FariaJunior_MnTe_EffHamA},
large anisotropic magnetoresistance and planer Hall effect~\cite{Kriegner_2016, Kriegner_2017, YinGen_Theory_MnTe, Wang_2023b, GonzalezBetancourt_2024},
and intrinsic anomalous Hall effect~\cite{Gonzalez_2023, Kluczyk_2024}.
These diverse properties of MnTe have attracted significant attention in both fundamental physics and device applications.
However, the connection between these spintronic phenomena and the physical mechanism is unclear.
The main reason is that in the Brillouin zone of the hexagonal $\alpha$-MnTe, the valence bands at the $\Gamma$ and $A$ point display quite distinct behaviors and
are both very close to the valence band maximum or the Fermi level~\cite{Krause_2013, YinGen_Theory_MnTe}.
Therefore, the transport properties, mainly determined by the electrons near the Fermi level, can be completely different under different conditions.  

In this article, we systematically analyzed the band structures of $\alpha$-MnTe near the Fermi level at the $\Gamma$ point and the $A$ point respectively 
based on the first-principles band structures, analytical tight-binding models, and group theory discussion. 
The analytical $k\cdot p$ effective Hamiltonians with SOC at the $\Gamma$ point and the $A$ point were obtained under two different in-plane N\'{e}el vector directions $[11\bar{2}0]$ and $[1\bar{1}00]$ respectively.
The derived effective Hamiltonian fits well to the first-principles band structures, 
and all the independent parameters indicated by the group theory were determined.
The strong anisotropic Fermi surfaces at the $\Gamma$ point originated from the SOC are distinct with different N\'{e}el vector orientations. 

\section{Methods}

\subsection{First principles calculations}
The electronic band structures of MnTe were obtained based on density-functional theory (DFT) based calculations with
projector augmented wave (PAW) pseudopotentials~\cite{Bloechl_1994, Kresse_1999},
implemented in the Vienna ab initio simulation (VASP) package~\cite{Kresse_1996, Kresse_1996a}.
The generalized gradient approximation in Perdew, Burke, and Ernzerhof
formation~\cite{Perdew_1996} was used as the exchange-correlation energy.
We employed the Hubbard $U$ method in the Liechtenstein implementation~\cite{Liechtenstein_1995} of $U=4.0\eV$, $J=0.9\eV$ 
on Mn($3d$) orbitals to include the on-site strong-correlation effects of the localized $3d$ electrons.
An energy cutoff of 600 eV was used for the plane-wave expansion
throughout the calculations.
The $\Gamma$-centered $12\times12\times8$ $k$-mesh are sampled in the Brillouin zone for self consistent calculations. 
The experimental lattice parameters of $a=4.171$\AA~ and $c=6.686$\AA~\cite{Kriegner_2016} are used.
The SOC was included when we considered the AFM phase with the N\'{e}el vectors along $x$ and $y$ directions

After we obtained the eigenstates and eigenvalues,
a unitary transformation from the plane-wave basis to the localized Wannier function (WF) basis 
was performed to construct the tight-binding Hamiltonian 
by using the band disentanglement method~\cite{Souza_2001}
implemented in the Wannier90 package~\cite{Mostofi_2014}.
Only Te($5p$) orbitals are chosen for the projection.

\subsection{General group representation theory}

To understand all valence bands and facilitate future transport study, 
the effective Hamiltonian of MnTe will be constructed using the representation theory of space group, spin group, and magnetic group. 
Only Te($5p$) orbitals are taken into account. 
Each unit cell contains two Te atoms. The Hilbert space is thus expanded by the basis of atomic orbitals $\phi^l$ from
A, B sublattices:
\begin{equation}
\psi_{A,B}^{\bk,l}(\br) =\sum_{j}e^{i\bk\cdot\bR^{j}}\phi^l(\br-\bR^{j}-\tau_{A,B})
\end{equation}
where $\bR^j$ are Bravais lattice vectors and $\tau_\alpha$ with $\alpha={A,B}$ are basis of two sublattices. For MnTe, $
\tau_{A}=\frac{2}{3}\mathbf{a}_{1}+\frac{1}{3}\mathbf{a}_{2}-\frac{1}{4}\mathbf{a}_{3}
$ and $\tau_{B}=\frac{1}{3}\mathbf{a}_{1}+\frac{2}{3}\mathbf{a}_{2}+\frac{1}{4}\mathbf{a}_{3}$.
Under a generic symmetry operation $\{g|\tau\}$, a lattice point from one unit cell could be transferred to another, so that $\{g|\tau\}\tau_{\alpha}=g\tau_{\alpha}+\tau=\tau_{\beta}+\bR'
$. To ensure $\{g|\tau\}|\psi_{\alpha}^{\bk,l}\ra$ is a linear superposition of $|\psi_{\alpha}^{\bk,l}\ra$ with the same $\bk$, $g$ must belong to the little group of $\bk$, i.e. $g\bk=\bk+\mathbf{K}$, where $\mathbf{K}$ is a lattice vector of the reciprocal space. It can be shown that $
\{g|\tau\}|\psi_{\alpha}^{\bk,l}\ra =\mathcal{D}_\gamma(g)_{ll'}e^{-i\bk\cdot\bR'}|\psi_{\beta}^{\bk,l'}\ra$,
where $\mathcal{D}_\gamma(g)$ is the irreducible representation $\gamma$ of $g$ in the atomic basis. Therefore, the representation of $\{g|\tau\}$ is
\begin{equation}
    D(\{g|\tau\})_{\alpha l,\beta l'}=\mathcal{D}_\gamma(g)_{ll'}e^{-i\bk\cdot\bR'},
\end{equation}
where the factor $e^{-i\bk\cdot\bR'}$ is important in the discussion of nonsymmorphic crystals away from the $\Gamma$ point.

A Hamiltonian $H=\sum_{\bk}h(\bk)|\psi^{\bk}\ra\la\psi^{\bk}|$ must be
invariant under a symmetry operator $g$ such that $\la g\psi^{\bk}|h(g\bk)|g\psi^{\bk}\ra=\la\psi^{\bk}|h(\bk)|\psi^{\bk}\ra$.
Given the representation of symmetry group under the basis $\{\psi^{\bk}\ra\}$:
$|g\psi^{\bk}\ra=D(g)|\psi^{\bk}\ra$, the invariance indicates $D(g)h(g^{-1}\bk)D^{-1}(g)=h(\bk)$. Due to the group representation homomorphism, only generators of the little group need to be considered to construct the effective Hamiltonian $h(\bk)$. If both the matrices $h_i$ and $f_i(\bk)$- a polynomial of $\bk$- expand the same irreducible representation of the little group, their direct product must contain the trivial irreducible representation satisfied by the Hamiltonian. Therefore one can construct the effective Hamiltonian $h(\bk)=\sum_{i=1}^d h_d f_d(\bk)$, where $d$ is the dimension of the representation.

Once symmetry operations include the time-reversal operation $\mathcal{T}$ that flips the crystal momentum, the condition for invariance is different. For a generic symmetry operation $g=g_{0}\mathcal{T}$, where $g_0$ is the unitary component, its representation of can be written as $D(g)=D(g_{0}\mathcal{T})K$. Here, $D$ is unitary and $K$ is a complex conjugate. Invariance of the Hamiltonian requires $D(g_{0}\mathcal{T})H^{*}(-g_{0}^{-1}\bk)D^{-1}(g_{0}\mathcal{T})=H(\bk)$ instead. Later we will show that the AFM state with N\'{e}el vector along $[1\bar{1}00]$ is exactly this case, while the N\'{e}el vector along $[11\bar{2}0]$ belongs to the simple case in the above paragraph.

\subsection{Tight binding theory}

The symmetric analysis provides all allowed terms in the effective Hamiltonian. However, some terms might vanish due to accidental degeneracy and some terms are tiny due to weak spin-orbit coupling perturbation. To confirm the band structure with orbital basis predicted by the symmetry analysis and select relevant terms, a tight binding Hamiltonian is constructed (see Supplementary Materials for details). The full $\bk$-dependent Hamiltonian is 
\begin{equation}
H^{\bk} = H_{\mathrm{Te-Te}}^{\bk} + H_{\mathrm{Te-Mn}}^{\bk} + H_{\mathrm{Mn-Mn}}^{\bk} + H_{\mathrm{SOC}}
\end{equation}
where $H_{\mathrm{Te-Te}}^{\bk}$ is the hopping between all three $p$ orbitals of two sublattices of Te atoms. 
Up to third neighbor hopping are taken into account, and the hopping integrals are determined by the Slater-Koster rules~\cite{Slater_LCAO, Konschuh_2010}. 
$H_{\mathrm{Te-Mn}}^{\bk}$ is the nearest neighboring hopping between $p-d$ orbitals. 
Subjected to the trigonal crystal field symmetry, $d$-orbitals of Mn are split to three levels with degenerate $d_{xz}/d_{yz}$ orbitals, 
degenerate $d_{x^2-y^2}/d_{xy}$ orbitals, and a non-degenerate $z^2$ orbital. 
Without loss of symmetry, only $d_{z^2}$ is considered in the tight-binding model. 
Due to the local magnetic moment of Mn, majority and minority spins have different spin-preserving hopping amplitudes, but the Slater-Koster rule for each spin is still respected. In addition, $H_{\mathrm{Mn-Mn}}^{\bk}$ accounts only for the on-site energy of the $d_{z^2}$ orbital, and thus the dispersion within the $d_{z^2}$-orbital subspace is neglected.
$H_{\mathrm{SOC}}$ is the onsite SOC of Te atoms.

Integrating over the $d_{z^2}$ by the Schrieffer-Wolff transformation~\cite{Schrieffer_1966, Liu_2011}, 
an effective Hamiltonian solely in terms of $p$-electrons can be derived as
\begin{equation}
H_{p}^{\bk} = H_{\mathrm{Te-Te}}^{\bk} + \Delta{H_{\mathrm{Te-Te}}^{\bk (2)}} + H_{\mathrm{SOC}}
\end{equation}\label{eq:TB_Te-Mn}
Here $\Delta{H_{\mathrm{Te-Te}}^{\bk (2)}} = \frac{H_{\mathrm{Te-Mn}}^{\bk} H_{\mathrm{Te-Mn}}^{\bk \dagger}}{\epsilon_p - \epsilon_d}$, 
where $\varepsilon_p$ and $\varepsilon_d$ are the energies of Te $p$ and Mn $d$ levels respectively. 
Minimal Hamiltonian at the $\Gamma$ and $A$ points can be derived by further application of the Schrieffer-Wolff transformation $H^{\bk}_{\mathrm{eff}} = H_{0}^{\bk} + V^{\bk}\frac{1}{\varepsilon_{0} - H^{'\bk}} V^{\bk \dagger}$, where $H^{'\bk}$ is the Hamiltonian of the Hilbert space to be integrated out. 
In the previous transformation $\Delta{H_{\mathrm{Te-Te}}^{\bk (2)}}$ above, $d_{z^2}-d_{z^2}$ hopping is neglected so that $d_{z^2}$ is dispersionless and $H^{'\bk} = H_{\mathrm{Mn-Mn}}^{\bk} = \varepsilon_d$. However, now $p$-orbitals to be integrated out are dispersive, so $H^{'\bk}$ has a matrix form. Here we write $\frac{1}{\varepsilon_{0} - H^{'\bk}}$ as $\frac{1}{(\varepsilon_0-\Lambda)-(H^{'\bk}-\Lambda)}$ and Taylor expand $(H^{'\bk}-\Lambda)$, 
where $\Lambda$ is the diagonal part of $H^{'\bk}$. 
We found that in order to derive an effective Hamiltonian with the correct symmetry, 
higher order terms of $(H^{'\bk}-\Lambda)$ must be kept. 
Details of the tight-binding theory can be found in the Supplementary Materials.

\section{First principles Band Structures}

\begin{figure}
\includegraphics[width=1\columnwidth]{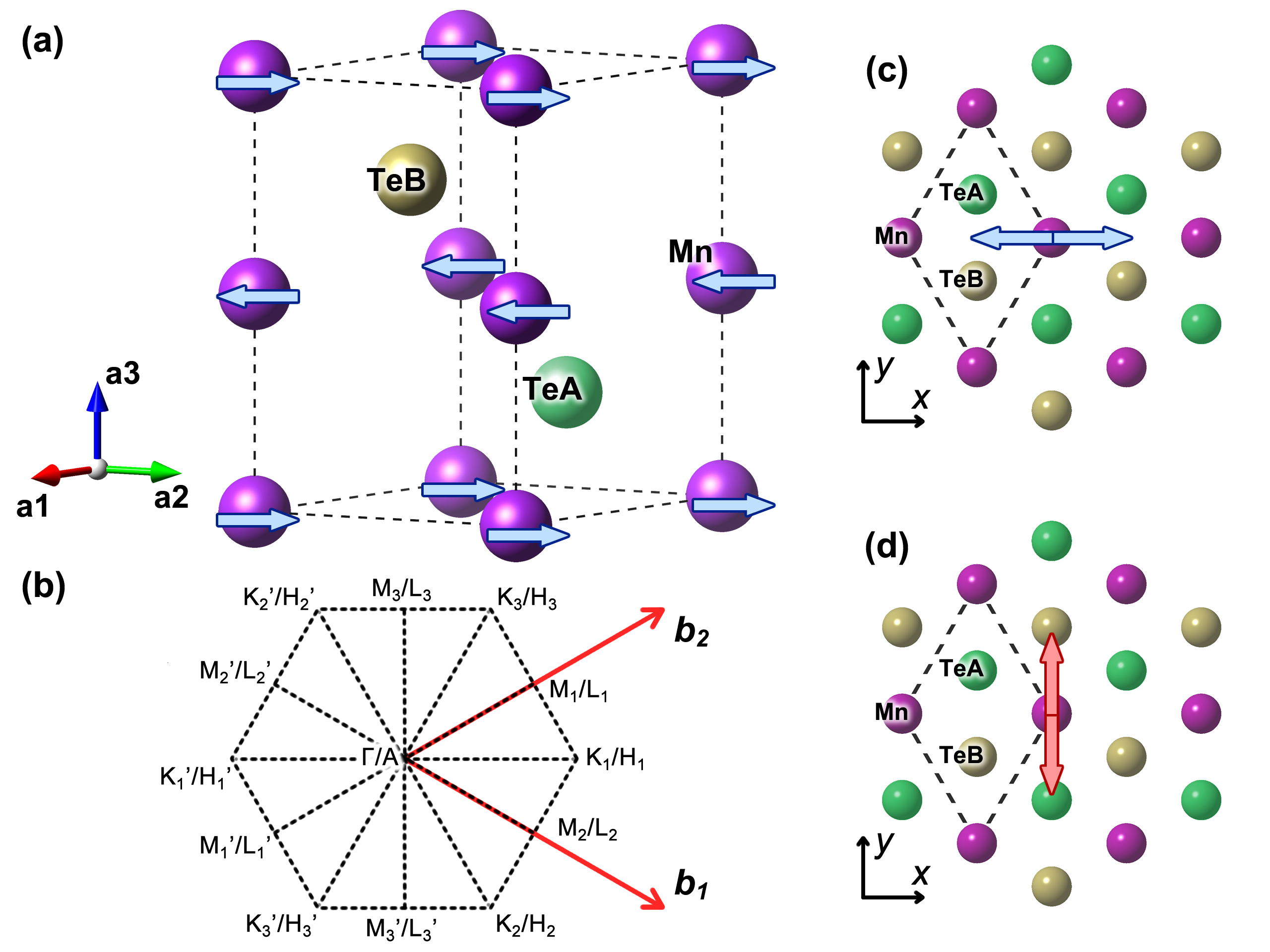}
\caption{The lattice structure of $\alpha$-MnTe under (a) the main view and (c)(d) top views. (b) the Brillouin zone of MnTe with all high symmetric $k$-points. Purple balls represent Mn atoms. Green and yellow balls represent Te atoms in A and B sublattices respectively. In (a), the local magnetic moments on Mn with layered antiferromagnetic ordering lay in the $xy$-plane. In (c) and (d), the N\'{e}el vectors are along $x$ and $y$ directions respectively. }
\label{fig:structure} 
\end{figure}

$\alpha$-MnTe has NiAs-type lattice structure. 
It is a nonsymmorphic space group $G=P_{6_{3/mmc}}$ (No. 194). 
The lattice structure and the Brillouin zone of MnTe are shown in Fig.~\ref{fig:structure}(a,b). 
Both Mn and Te form hexagonal lattices in the plane. 
A-type collinear antiferromagnetic order is built by Mn. 
Easy plane anisotropy is suggested experimentally~\cite{Kriegner_2016, Kriegner_2017}, 
and the anisotropy in the plane is negligibly small ($< 0.05\meV$ per unit cell). 
Each local moment on the magnetic Mn atom thus can point along $[11\bar{2}0]$ direction (the $x$ direction) (Fig.~\ref{fig:structure}(c)) or $[1\bar{1}00]$ direction (the $y$ direction) (Fig.~\ref{fig:structure}(d)) with close probability. 
Due to the opposite direction of Mn's magnetization in neighboring layers, two Te atoms, labeled as TeA and TeB, and marked by green and brown respectively to be distinguished from each other. 

\begin{figure}
\includegraphics[width=1\columnwidth]{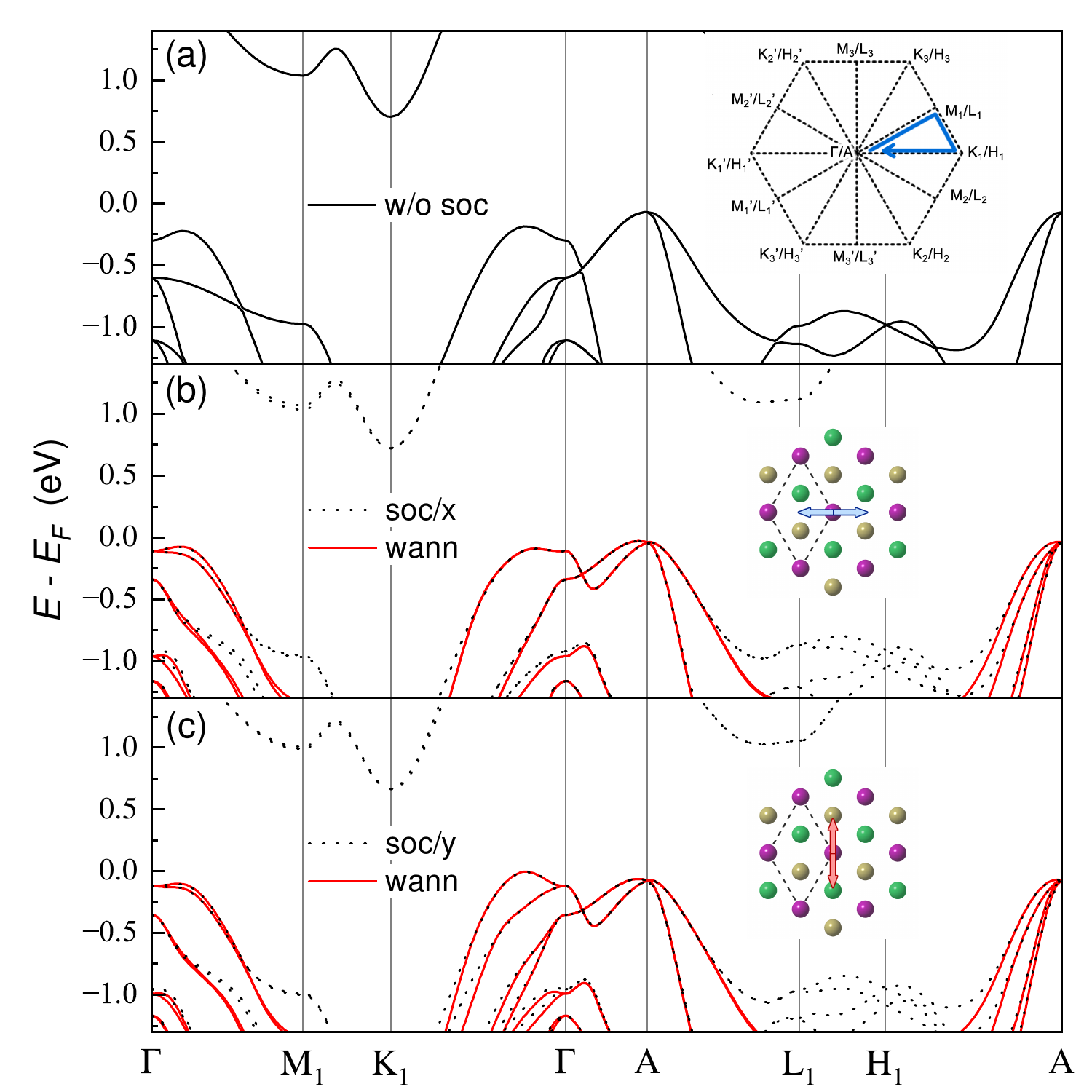}
\caption{The band structures of A-type antiferromagnetic MnTe. 
(a) without spin-orbit coupling (SOC),
with SOC included and with the N\'{e}el vector aligned along 
(b) the $x$ direction and (c) the $y$ direction respectively.
Insert in (a) shows the $k$-point path in a hexagonal Brillouin zone for the plotting band structure.
The Fermi energy, which is the valence band maximum (VBM), is set to zero. 
}
\label{fig:bands} 
\end{figure}

The band structures for MnTe are shown in Fig.~\ref{fig:bands}. 
The band gap is about 0.7 eV, indicating the semiconducting behavior.
The WF-based Hamiltonian has the same eigenvalues as
those obtained by first-principles calculations
from 0.5 eV below the Fermi energy to the valence band maximum (VBM). 
This also indicates that electrons from Mn are away from the Fermi energy 
so that only $p$-orbitals of the Te electrons need to be taken into account
when we study spintronic properties.
The $\Gamma$ and $A$ points are both close to VBM.
In addition, bands with the N\'{e}el vector aligned along the $x$ and $y$ directions 
have distinct behaviors when SOC is included, shown in Fig.~\ref{fig:bands}(b)(c).
We should separately discuss them in detail.

\section{Results of Effective Hamiltonian}


\subsection{Non-SOC paramagnetic phase}

The quotient group $G/T$ of MnTe's space group $G$ with respect to its translation group $T$ is isomorphic to $D_{6h}$. 
It has 24 symmetry operations with the generators
$\{3_{0001}^{-}|0\}$, $\{2_{0001}|1/2\}$, $\{2_{110}|0\}$, and $\mathcal{I}=\{-1|0\}$. According to the orbital-resolved band structure plot in Fig.~\ref{fig:nonsoc}(a), the valence band around the $\Gamma$-point at the Fermi surface is dominated by the single $p_z$ orbital.
In this paramagnetic state, real spin $s$ is irrelevant, so that sub-Hilbert space is two dimensional expanded by $A/B$ pseudospin, which is labeled as $\sigma$. In this representation,
these generators are $\{3_{0001}^{-}|0\}=\sigma_0$, $\{2_{0001}|1/2\}=\sigma_x$, $\{2_{11\bar{2}0}|0\}=-\sigma_x$, and $\mathcal{I}=-\sigma_x$.

%

Representation is different at the point $A=(0,0,1/2)$ which sits right on top of the $\Gamma$ point. The little group of $A$ is still $D_{6h}$, so they share the same generators. But the fractal translations in this nonsymmorphic group matters away from the $\Gamma$ point. Furthermore, the DOS around the $A$ point [Fig.~\ref{fig:nonsoc}(a)] shows dominance from the $p_x$ and $p_y$ orbitals. Using $\tau$ to represent the $p_x/p_y$ pseudospin, representations of these generators are $\{3_{0001}^{-}|0\}=\sigma_{0}(-\frac{1}{2}\tau_{0}+\frac{\sqrt{3}}{2}i\tau_{y})$, $\{2_{0001}|1/2\}=i\sigma_{y}\tau_{0}$, $\{2_{11\bar{2}0}|0\}=\sigma_x\tau_z$, and $\mathcal{I}=-\sigma_x$.

The band structure of the paramagnetic state without spin-orbit coupling can be derived. The full table irreducible representation is in the Table~I of the Supplementary Materials. It is, however, worthwhile to discuss the one dimensional trivial representation $A_{1g}$: $D_{A_{1g}}(g)=1$ for any operation $g$. At the $\Gamma$ point, other than the trivial representation $\sigma_0s_0$, another irreducible representation is $\sigma_x$. The four fold degeneracy of $|p_{zA}\uparrow\ra, |p_{zB}\uparrow\ra, |p_{zA}\downarrow\ra,|p_{zB}\downarrow\ra$ is thus lifted, with each identified by the eigenvalues $\pm1$ of $\sigma_x$. These correspond to the symmetrized ($\ket{S}$) and anti-symmetrized ($\ket{A}$) states between $A$ and $B$ sublattices; $\ket{S(A)}=p_{zA}\pm p_{zB}$ . Each band has two-fold spin degeneracy. This is consistent with the DFT calculations. The valence band top at the $\Gamma$ point are the spin-degenerate anti-bonding states with the symmetrized $\ket{S}$. In contrast, only the identity matrix $\sigma_0\tau_0 s_0$ expands the trivial irreducible representation at the A point. As a result, all eight bands are degenerate in the paramagnetic state without spin-orbit coupling. Using the full irreducible representation Table.~I in the Supplementary Materials, one can easily construct the full non-magnetic Hamiltonian of MnTe effective around the $\Gamma$ and $A$ points, but they are not the main goal here.

\subsection{Non-SOC AFM phase}

The band structure changes when the A-type antiferromagnetic order is introduced. Without spin-orbit coupling, spin and the orbital space can be rotated independently, so that this magnetic state satisfies spin point group $^26/^2m^2m^1m$. This group is isomorphic to $D_{6h}$ group, but the
generators are $\{3_{0001}^{-}|0\}$, $\{2_{11\bar{2}0}|0\}$, $\{2_{0001}|1/2\}\mathcal{R_{\pi}}$,
and $\mathcal{I}$. Here $\mathcal{R}_{\pi}$ is the spin flip operator. At the absence of spin-orbit coupling, one has the liberty of choosing arbitrary spin quantization axis and the physics is unchanged. For simplicity, let's use $z$ as the spin quantization axis of Mn layer, then $\mathcal{R}_{\pi}$
can be either $\mathcal{R}_{1}=e^{-i\pi s_{x}/2}=-is_{x}$
or $\mathcal{R}_{2}=e^{-i\pi s_{y}/2}=-is_{y}$. Its irreducible representations are listed in Table \ref{Table:representation_spin_group_nonsoc}.

\begin{table*}[t]
\begin{centering}
\begin{tabular}{|c|c|c|c|c|c|c|c|c|}
\hline 
 & $E$ & $\{3_{0001}^{-}|0\}$ & $\{2_{0001}|1/2\}\mathcal{R}_{\pi}$ & $\{2_{11\bar{2}0}|0\}$ & $\mathcal{I}$ &  & $\Gamma$ & $A$\tabularnewline
\hline 
\hline 
$A_{1g}$ & 1 & 1 & 1 & 1 & 1 & $x^{2}+y^{2}$, $z^{2}$ & $\sigma_{0}$, $\sigma_{x}$ & $\sigma_{0}\tau_{0}$, $\sigma_{x}\tau_{0}s_{z}$\tabularnewline
\hline 
$A_{2u}$ & 1 & 1 & 1 & -1 & -1 & $z$ & $\sigma_{y}s_{z}$, $\sigma_{z}s_{z}$ & $\sigma_{y}\tau_{0}$, $\sigma_{z}\tau_{0}s_{z}$\tabularnewline
\hline 
$E_{2g}$ & 2 & -1 & 2 & 0 & 2 & $(x^{2}-y^{2},2xy)$ &  & $\sigma_{0}(\tau_{z},\tau_{x})$, $\sigma_{x}(\tau_{z},\tau_{x})s_{z}$\tabularnewline
\hline 
$E_{1g}$ & 2 & -1 & -2 & 0 & 2 & $(yz,xz)$ &  & $\sigma_{x}(\tau_{z},\tau_{x})$, $\sigma_{0}(\tau_{z},\tau_{x})s_{z}$\tabularnewline
\hline 
$E_{1u}$ & 2 & -1 & -2 & 0 & -2 & $(x,y)$ &  & $\sigma_{y}(\tau_{x},\tau_{z})s_{z}$, $\sigma_{z}(\tau_{x},\tau_{z})s_{z}$\tabularnewline
\hline 
\end{tabular}
\par\end{centering}
\caption{Irreducible representations of spin group $^26/^2m^2m^1m$, corresponding to AFM phase without SOC included.}
\label{Table:representation_spin_group_nonsoc}
\end{table*}

\begin{figure}
\includegraphics[width=1\columnwidth]{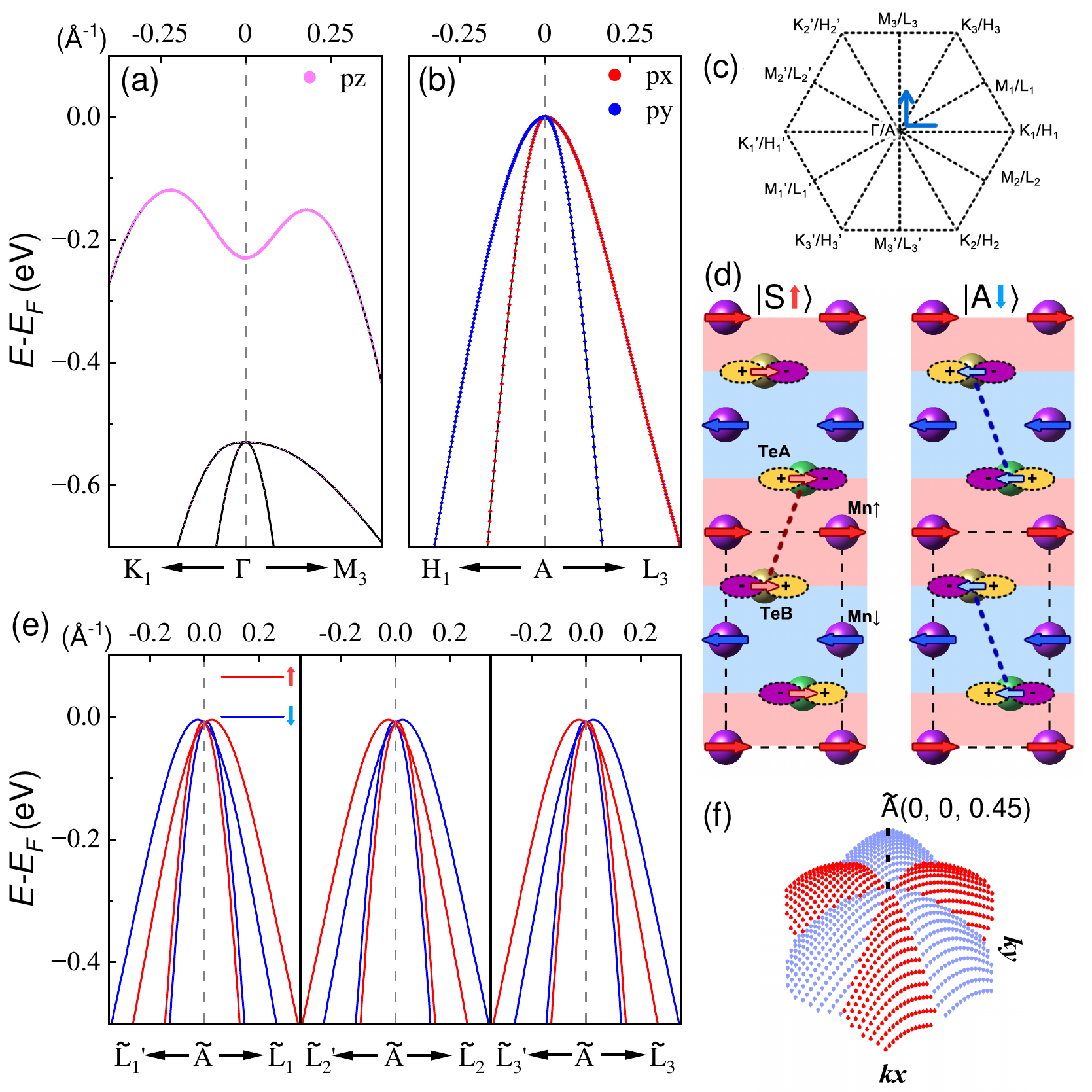}
\caption{The detailed band structure of the valence bands without SOC.
(a)(b)
The band structures with Te($5p$) orbital component near (a) the $\Gamma$ point and (b) the $A$ point, respectively. 
The $k$-point path is shown in (c).
(d) Two degenerate states at $A$ point, 
the symmetrized $A-B$ sublattices with up spin $\ket{S\uparrow}$ 
and the anti-symmetrized $A-B$ sublattices with down spin $\ket{A\downarrow}$.
(e) the band structure in the $k_x k_y$ plane where $kz = 0.45 \frac{2\pi}{c}$ (near $\tilde{A} = (0, 0, 0.45)$) in the reciprocal space.
Red and blue curves represent spin up and down bands respectively.
(f) the energy surface for the valence band in $kx$-$ky$ plane near $\tilde{A}$ point. 
The red and blue  dots represent the spin up and down respectively. 
The units of k-points is \AA$^{-1}$.}

\label{fig:nonsoc} 
\end{figure}

At the $\Gamma$ point, according to the discussion above, anti-bonding states with eigenstate $-1$ of $\sigma_x$ are split from bonding states; the latter
are far from the Fermi level and many other Mn and Te bands are sandwiched in between. 
We are thus only interested in the minimal Hamiltonian in the two-dimensional sub-Hilbert space expanded by anti-bonding states $\{\ket{S\uparrow}, \ket{S\downarrow}\}$. 
This is consistent with the result obtained by the tight binding theory in Supplementary Materials.
Projection to the corresponding eigenstates, 
$\sigma_ys_z$ and $\sigma_zs_z$ both vanish since they are relevant only in the transition between bonding and anti-bonding states. 
As a result, only trivial irreducible representation $A_{1g}$ is relevant and the Hamiltonian of this valence state is given by
\begin{align}
H_{\mathrm{non}\mathchar`- \mathrm{SOC}}^{\Gamma} &= c_{1} \left(k_{x}^{2}+k_{y}^{2}\right) + c_{2} k_{z}^{2} + c_{3} \left(k_{x}^{2} + k_{y}^{2}\right)^{2} \nonumber \\
&+ c_{4} \left[4 \left(k_x^2 - k_y^2\right)^{3} - 3 \left(k_x^{2} - k_y^{2}\right) \left(k_x^{2} + k_y^{2}\right)^{2}\right]
\label{eq:gamma_nonsoc}
\end{align}
Here, terms $c_{1,2,3}$ are directly constructed from Table \ref{Table:representation_spin_group_nonsoc}. 
Although immediately next to the $\Gamma$ point, the DFT results in Fig.~\ref{fig:bands}(a) can be captured by these three terms, the band structure shows non-quadratic non-monotonous band along the $\Gamma-K/M$ line away from the $\Gamma$ point. 
This is an important feature since the real Fermi surface might cut the valence band top away from the $\Gamma$ point. 
Along $\Gamma-K/M$ line the little group symmetry is much reduced than the $\Gamma$ point, 
so the higher order $\bk$ corrections cannot rely on the irreducible representation in Table.~\ref{Table:representation_spin_group_nonsoc}. 
We notice that on the $\Gamma-K-M$ plane the band exhibit a 6-fold rotational symmetry. 
Therefore $c_4$ term is included, which is identical to $k^6\sin^6\theta \cos6\phi$ in the spherical coordinate.
This is a spinless Hamiltonian, meaning the band is degenerate. 
This fact is consistent with the band structure in Fig.~\ref{fig:bands}(a). 
Along any direction from the $\Gamma$ point, bands are always spin degenerate. 
If altermagnet is narrowly defined as the spin splitting in the absence of spin-orbit coupling, MnTe is \textit{not} an altermagnet, at least from the transport perspective once the $\Gamma$ point is dominant.

At $A$ point, $\sigma_{x}\tau_{0}s_{z}$ term of $A_{1g}$ irreducible representation splits the 8-fold degenerate
bands into two 4-fold degenerate manifolds, 
depending on the eigenvalues of this matrix. 
$p_x$ and $p_y$ pseudospins are always degenerate due to the presence of $\{3^-_{0001}|0\}$ operation.  
The anti-symmetrized $A-B$ sublattices with up spin ($\ket{A\uparrow}$) and the symmetrized $A-B$ sublattices with down spin ($\ket{S\downarrow}$) are degenerate.  
On the other hand, 
symmetrized $A-B$ sublattices with up spin ($\ket{S\uparrow}$) and the anti-symmetrized $A-B$ sublattices with down spin ($\ket{A\downarrow}$) are also degenerate, 
but significantly split from the former. 
The latter is close to the Fermi energy as indicated by the DFT result.
Under A-type antiferromagnetic ordering, the mixture of sublattice and spin degrees of freedom is a key feature of bands at the $A$ point 
due to alternating phases of $p$-orbitals along $\left[0001\right]$ direction as shown in Fig.~\ref{fig:nonsoc}(d). 
The symmetrized $A-B$ sublattices within the unit cell is identical to the anti-symmetrized state between Te atoms from two neighboring unit cells along $\left[0001\right]$. 
However the spin state of these two equivalent interpretations are opposite due to alternating Mn spins along $\left[0001\right]$. 
Let's use a new pseudospin $\mathbf{\omega}$ to indicate this two-fold degeneracy between $\ket{S\uparrow}$ and $\ket{A\downarrow}$. 
Although not exactly the real spin operator, it is still a pseudo vector like the real spin. 
Minimal Hamiltonian is expanded by four-dimensional matrices $\omega_i\tau_j$. 
Projecting onto this sub-Hilbert space, only terms commuting
with $\sigma_{z}\tau_{0}s_{z}$ could stay. 
All other terms enable inter-band transition to the cluster of bands far below the valance band top. 
The resulting effective Hamiltonian is therefore
\begin{align}
H_{\mathrm{non}\mathchar`- \mathrm{SOC}}^{A} &= c_{1} \left(k_{x}^{2}+k_{y}^{2}\right) + c_{2} k_{z}^{2} + c_{3} \left[\left(k_{x}^{2} - k_{y}^{2}\right) \tau_{z} + 2 k_{x} k_{y} \tau_{x}\right] \nonumber \\ 
&+ c_{4} \omega_{z} \left(k_{y} k_{z} \tau_{z} + k_{x} k_{z} \tau_{x}\right) 
\label{eq:A_nonsoc}
\end{align}
This effective Hamiltonian is consistent with the DFT result. 
At $A-\Gamma$ line, $k_{x}=k_{y}=0$ so no splitting is expected. 
As shown in Fig.~\ref{fig:nonsoc}(b),
at $A-L_3$ line, $k_{x}=k_{z}=0$,
the Hamiltonian is reduced to $H_{\mathrm{non}\mathchar`- \mathrm{SOC}}^{A-L_3}=c_{1}k_{y}^{2}-c_{3}k_{y}^{2}\tau_{z}$, 
so that splitting between $p_{x}$ and $p_{y}$ is observed. 
The same is expected
along $A-H_1$ line with $k_{y}=k_{z}=0$, where $H_{\mathrm{non}\mathchar`- \mathrm{SOC}}^{A-H_1}=c_{1}k_{x}^{2}+c_{3}k_{x}^{2}\tau_{z}$. 
Along all these high symmetry lines, spin splitting is absent due to the irrelevance of $\omega$ in the Hamiltonian.
DFT shows bulk $g-$wave spin splitting only at low symmetry areas in the Brillouin zone [Fig.~\ref{fig:nonsoc}(e)(f)].
One should note that the spin-dependent second ordered $c_{4}$ term is not enough to describe the bulk $g-$wave spin splitting with three-fold symmetry.
Since the spin splitting next to $A$ point does not push the bands above the energy of the $A$ point apparently, 
it may not impact transport properties effectively.
In addition, SOC and the corresponding in-plane magnetic anisotropy can reduce the symmetry and the high-order terms are not as significant.
Therefore, we did not further introduce higher-order terms here.

From the tight-binding theory, we analytically derived all the terms permitted by the symmetry constraints in Eqs.~\ref{eq:gamma_nonsoc} and \ref{eq:A_nonsoc} (see Supplementary Materials for details). In the absence of SOC, the valence band maximum at the $\Gamma$ point is an anti-bonding state between the A and B sublattices and remains spin-degenerate. At the $A$ point, the states become degenerate, corresponding to $\ket{S\uparrow}$ and $\ket{A\downarrow}$ states of both $p_{x}$ and $p_{y}$ orbitals. These findings are all consistent with the symmetry analysis, thereby validating our tight-binding model.

\subsection{SOC AFM with N\'{e}el vector along $[11\bar{2}0]$}

\begin{table*}
\begin{tabular}{|c|c|c|c|c|c|c|c|}
\hline 
 & $E$ & $\{2_{0001}|1/2\}$ & $\{2_{11\bar{2}0}|0\}$ & $\mathcal{I}$ &  & $\Gamma$ & $A$\tabularnewline
\hline 
\hline 
$A_{g}$ & 1 & 1 & 1 & 1 & $x^{2}$, $y^{2}$, $z^{2}$, const. & $\sigma_{0}$, $\sigma_{x}$ & $\sigma_{0}\tau_{0,z}$, $\sigma_{x}\tau_{0,z}s_{x}$\tabularnewline
\hline 
$A_{u}$ & 1 & 1 & 1 & -1 &  &  & $\sigma_{y}\tau_{0,z}s_{z}$, $\sigma_{z}\tau_{0,z}s_{y}$\tabularnewline
\hline 
$B_{1g}$ & 1 & 1 & -1 & 1 & $xy$ & $\sigma_{0}s_{z}$, $\sigma_{x}s_{z}$ & $\tau_{x,y}$, $\sigma_{x}\tau_{x,y}s_{x}$\tabularnewline
\hline 
$B_{1u}$ & 1 & 1 & -1 & -1 & $z$ &  & $\sigma_{y}\tau_{x,y}s_{z}$, $\sigma_{z}\tau_{x,y}s_{y}$\tabularnewline
\hline 
$B_{3g}$ & 1 & -1 & 1 & 1 & $yz$ & $\sigma_{0}s_{x}$, $\sigma_{x}s_{x}$ & $\sigma_{x}\tau_{0,z}$, $\sigma_{0}\tau_{0,z}s_{x}$\tabularnewline
\hline 
$B_{3u}$ & 1 & -1 & 1 & -1 & $x$ &  & $\sigma_{y}\tau_{0,z}s_{y}$, $\sigma_{z}\tau_{0,z}s_{z}$\tabularnewline
\hline 
$B_{2g}$ & 1 & -1 & -1 & 1 & $xz$ & $\sigma_{0}s_{y}$, $\sigma_{x}s_{y}$ & $\sigma_{x}\tau_{x,y}$, $\sigma_{0}\tau_{x,y}s_{x}$\tabularnewline
\hline 
$B_{2u}$ & 1 & -1 & -1 & -1 & $y$ &  & $\sigma_{y}\tau_{x,y}s_{y}$, $\sigma_{z}\tau_{x,y}s_{z}$\tabularnewline
\hline 
\end{tabular}
\caption{Irreducible representations of $D_{2h}$}
\label{Table:representation_100soc}
\end{table*}

\begin{figure}
\includegraphics[width=1\columnwidth]{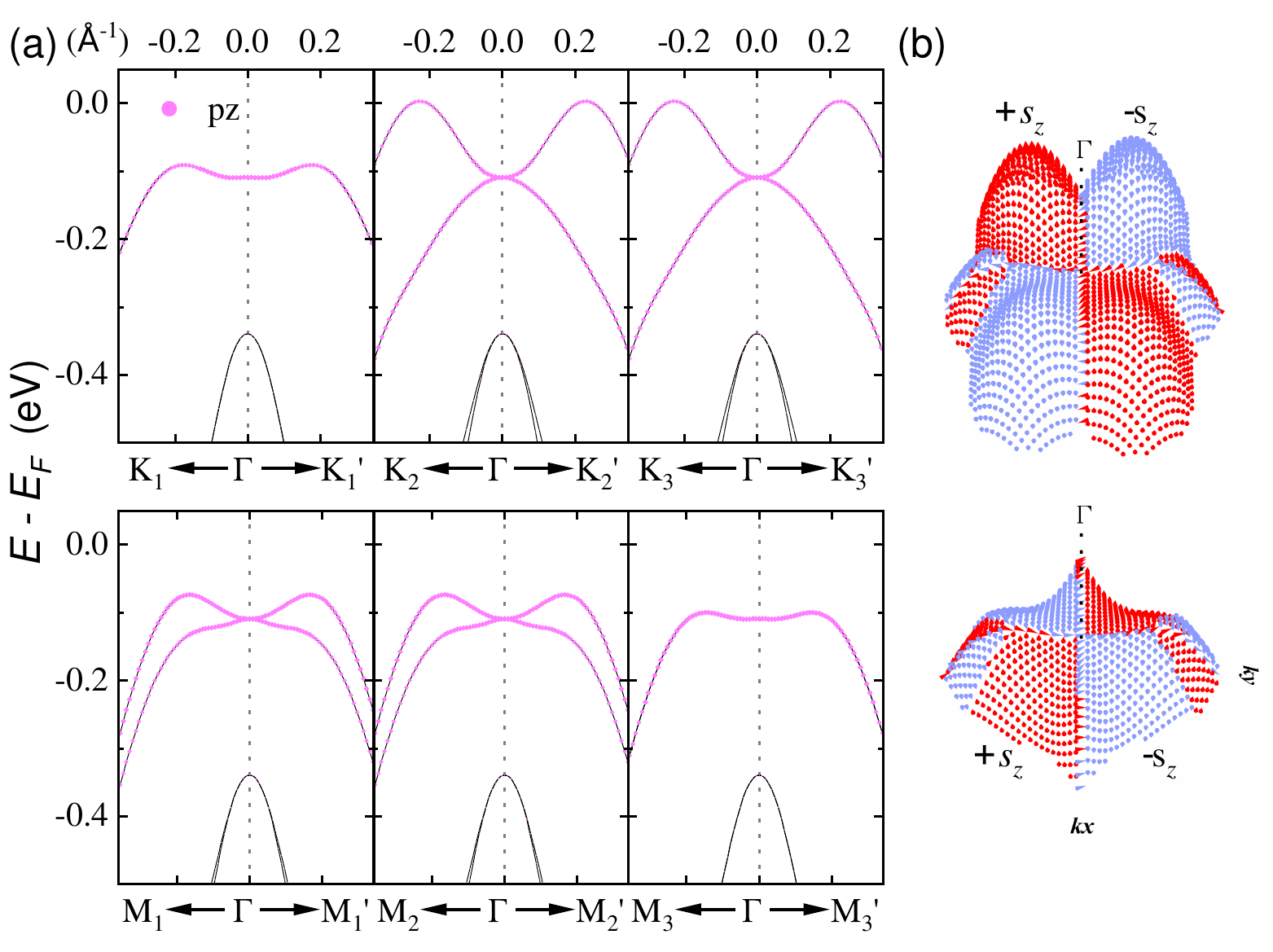}
\includegraphics[width=1\columnwidth]{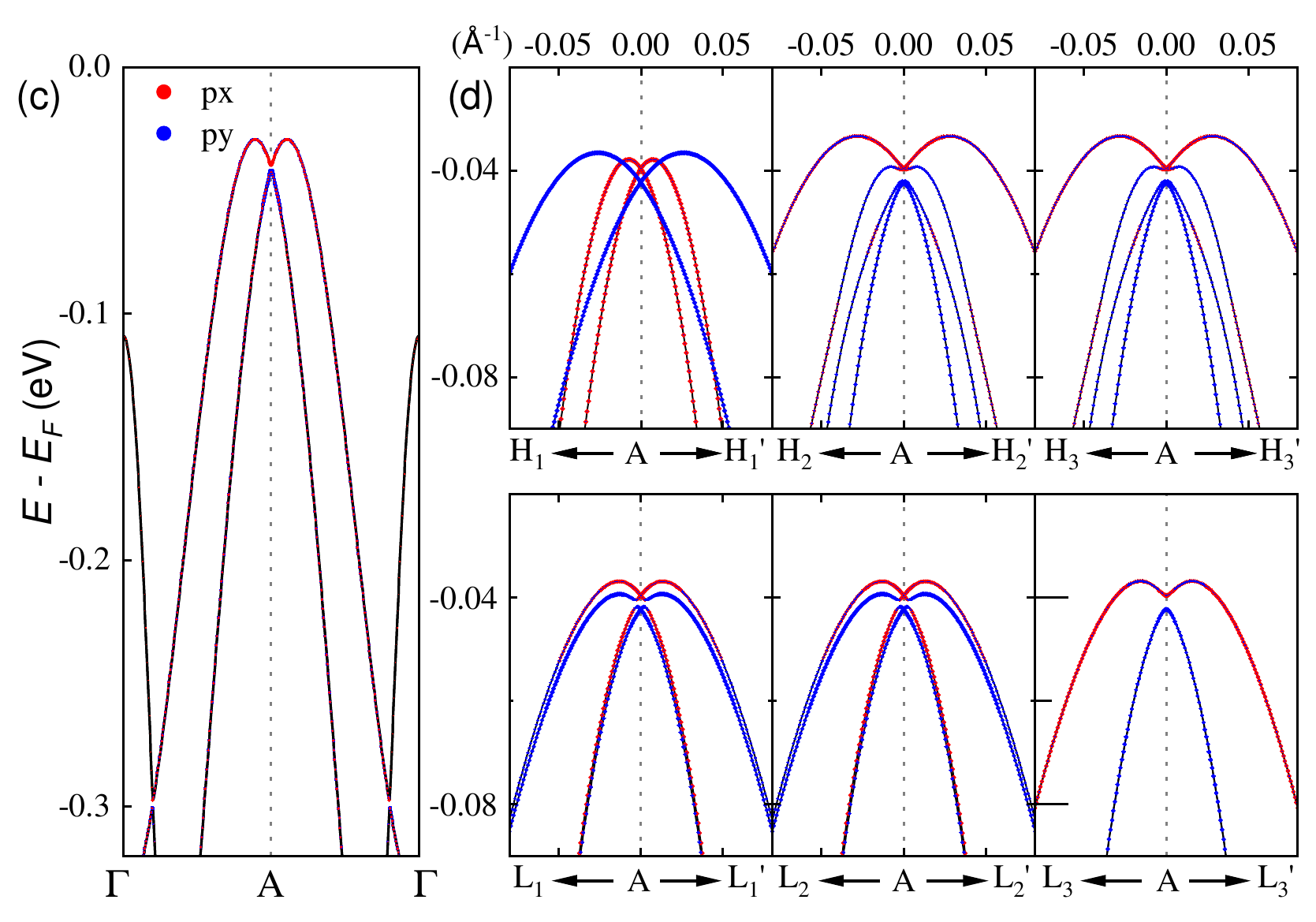}
\caption{
With the N\'{e}el vector aligned along the $[11\bar{2}0]$ ($x$) direction, 
The band structure with Te($5p$) orbital component (a) near the $\Gamma$ point, 
(c) along the $\Gamma$-A-$\Gamma$ $k$-point path, and (d) near A point. 
(b) the energy surface for the top two valence bands in $kx$-$ky$ plane near the $\Gamma$ point. 
The red and blue arrows on each dots represent the $+s_z$ and $-s_z$ spin directions respectively. 
The units of $k$-points is \AA$^{-1}$.}
\label{fig:af1x} 
\end{figure}

Significant spin splitting in MnTe is actually facilitated by the spin-orbit coupling. 
When the spin-orbit coupling is turned on, 
due to the co-rotation of both orbital and spin coordinates, 
the lattice symmetry is reduced and different antiferromagnetic orderings give distinct band structures. 
According to the DFT calculation, two antiferromagnetic orders with different in-plane directions of the N\'eel vectors should be considered.
When the N\'eel lies in $[11\bar{2}0]$
direction ($x$ direction in Fig.~\ref{fig:structure}(c-d)), the inversion $\mathcal{I}$ is still a symmetry,
but $C_{3}$ and $C_{6}$ symmetries are broken. 
The little groups at the $\Gamma$ and $A$ points are both isomorphic to $D_{2h}=D_{2}\otimes Z_{2}$,
where $Z_2=\{E=\{1|0\}, \mathcal{I}\}$, and $D_{2} = \{ \{2_{11\bar{2}0}|0\}, \{2_{0001}|1/2\}, \{2_{1\bar{1}00}|1/2\}, E=\{1|0\} \}$. 
Usually once the magnetic ordering is concerned, the time reversal operation $\mathcal{T}$ would appear in the space group. 
Interestingly, in the current case, $\mathcal{T}$ is not a symmetry operation even in combination with other unitary rotations. 
So the magnetic space group is the same as the conventional Fedorov space group. 
Absence of the time reversal symmetry thus naturally leads to the Hall effect in the absence of the external magnetic field or net magnetization~\cite{YinGen_Theory_MnTe}. 

At the $\Gamma$ point acting on the $p_z$ orbitals, the symmetry operations are $\{2_{0001}|1/2\}=\sigma_{x}s_{z}$,
$\{2_{11\bar{2}0}|0\}=\sigma_{x}s_{x}$, $\mathcal{I}=-\sigma_{x}$. While at the
$A$ point acting on the $p_{x,y}$ orbitals, $\{2_{0001}|1/2\}=i\sigma_{y}\tau_{0}s_{z}$, $\{2_{11\bar{2}0}|0\}=\sigma_{x}\tau_{z}s_{x}$,
and $\mathcal{I}=-\sigma_{x}$. The character table is given in Table \ref{Table:representation_100soc}.
For simplicity, only terms commute with $\sigma_x$ and $\sigma_{x}\tau_{0}s_{x}$ at the $\Gamma$ and $A$ points, respectively, are listed. These operations keep the Hilbert space
within the subspace given by the degenerate eigenstates required by their trivial irreducible representations. 

At the $\Gamma$ point, all remaining terms keep the anti-bonding state invariant, thus the $\sigma$-component can be neglected. 
Only the spin rotations are relevant. 
As a result, all allowed terms in addition to the non-SOC effective Hamiltonian are $k_{x} k_{y} s_{z}$, $k_{y} k_{z} s_{x}$, and $k_{x} k_{z} s_{y}$. 
They are contributions from the SOC. 
The tight-binding calculation indicates that only $k_{x} k_{y} s_{z}$ term is relevant when $k_{z}$ is tiny. 
The effective Hamiltonian is thus given by
\begin{align}
H_{[11\bar{2}0]}^{\Gamma} &= H_{\mathrm{non}\mathchar`- \mathrm{SOC}}^{\Gamma} + \gamma_{1} k_{x} k_{y}s_{z} + \left(\delta_{1} k_{x}^{3} k_y + \delta_{2} k_{y}^{3} k_{x} \right) s_{z} 
\label{eq:gamma_100}
\end{align}
Here as suggested in the authors' earlier work \cite{YinGen_Theory_MnTe}, 
a quartic SOC term $\left(\delta_{1} k_{x}^{3}k_y+\delta_{2}k_{y}^{3}k_{x}\right) s_{z}$ is included to enable the non-monotonous band around the $\Gamma$ point. 
This result is consistent with the spin resolved band from DFT calculation.
All terms with $s_{z}$ are combined with $k_{x} k_{y}$, so that 
the valence band around the $\Gamma$ point is spin polarized along $+\hat{z}$ direction in the II/IV quadrants and along $-\hat{z}$ in the I/III quadrants, as shown in Fig.~\ref{fig:af1x}(b).
The degenerate lines with zero is for all the $s_{z}$ terms are two lines $k_{x}=0$ and $k_{y}=0$.
Fig.~\ref{fig:af1x}(b) also shows that the sub valence band has the opposite spin direction.
Importantly, the tight binding calculation shows that $\gamma_1$ term is a first-order correction in SOC.
The large SOC of $0.5~\eV$ in Te thus explains the large spin splitting around the $\Gamma$ point. 
The energy maximum are located at four points symmetrically in four quadrants.
Considering its $p$-type semiconductor nature, the Fermi surface formed by the valence band forms four hole pockets which are mirror symmetric with each other and exhibits strong anisotropic characteristics as well as spin orientation, 
leading to a wealth of anisotropic transport properties.
This result has been well discussed in the authors' earlier work \cite{YinGen_Theory_MnTe}.

We employed the conjugate gradient method~\cite{CG_Fletcher_1964} to fit all the coefficients of the effective Hamiltonian obtained from our analysis. 
DFT bands in an ellipsoidal region of the Brillouin zone around the $\Gamma$ or $A$ point were chosen for the fitting. 
Considering that the SOC bands from DFT are obtained through self-consistent field calculations, 
the non-SOC potentials for different N\'{e}el vectors may vary slightly. 
Therefore, we performed separate fittings for the coefficients of the non-SOC part of the bands for different N\'{e}el vectors.
In particular, due to the significant impact of SOC on the bands and the presence of important spin-splitting features apart from the $\Gamma$ point, 
we adopted the transformation of $k_{x} \rightarrow \frac{1}{a} \sin k_{x}a, k_{y} \rightarrow \frac{1}{a} \sin k_{y}a$ to fit the coefficients of the part of SOC near the $\Gamma$ point.
The fitting parameters were obtained as 
$c_{1}/a^2=0.0306~\eV$, 
$c_{2}/c^2=-0.492~\eV$, 
$c_{3}/a^4=-0.0497~\eV$, 
$c_{4}/a^6=0.0047~\eV$, 
$\gamma_{1}/a^{2}=0.353~\eV$, 
$\delta_{1}/a^{4}=-0.381~\eV$, 
$\delta_{2}/a^{4}=0.0870~\eV$.

At the $A$ point, 
the same $\mathbf{\omega}$ matrices are used to represent the pseudospin of degenerate basis $\ket{S+}$ and $\ket{A-}$. Different as the non-SOC AFM case, here the spin space itself is no longer $SU(2)$ invariant. $\ket{+}$ and $\ket{-}$ states are determined by the eigenstates of $s_x$ instead, i.e., $s_x\ket{\pm}=\pm\ket{\pm}$.
Allowed independent SOC terms are 
$\tau_{z}$, 
$k_{x,y,z}^{2}\tau_{z}$, 
$k_{x}k_{y}\tau_{x,y}$, 
$k_{z}\tau_{x,y}\omega_{y}$, 
$k_{y}k_{z}\tau_{0,z}$, 
$k_{x}\tau_{0,z}\omega_{x}$, 
$k_{x}k_{z}\tau_{x,y}$, 
and $k_{y}\tau_{x,y}\omega_{x}$. 
The tight-binding calculation indicates the Hamiltonian given by
\begin{align}
   H_{[11\bar{2}0]}^{A} &= H_{\mathrm{non}\mathchar`- \mathrm{SOC}}^{A} + \gamma_{0} \tau_{z}  + \gamma_{1} \left(k_{x} \tau_{z} - k_{y} \tau_{x} \right) \omega_{x} + \gamma_{2} k_{y} \tau_{y}  \omega_{x} \nonumber \\
   &+ \gamma_{3} k_{z} \tau_{y} \omega_{y} + \gamma_{4} k_{x} k_{y} \tau_{y}  + \gamma_{5} k_{x} k_{z} \tau_{y} \omega_{z}     
\label{eq:A_100}
\end{align}
An interesting term is $\tau_z$ which opens up a gap at the $A$ point between $p_x$ and $p_y$ orbitals. 
That is a result of the broken $C_3$ symmetry. $p_x$ and $p_y$ orbitals are independent. 
But tight-binding shows that the gap $2\gamma_0$ is on the second order of SOC. 
DFT bands identify this gap as $3\meV$, shown in Fig.~\ref{fig:af1x}(d). 
It is thus negligibly small. 
The fitting parameters from the DFT band results were obtained as
$c_{1}/a^2=1.089~\eV$, 
$c_{2}/c^2=-0.129~\eV$, 
$c_{3}/a^2=-0.615~\eV$, 
$c_{4}/a^2=-0.0035~\eV$, 
$\gamma_{0}=0.0013~\eV$, 
$\gamma_{1}/a^{2}=-0.111~\eV$, 
$\gamma_{2}/a^{2}=0.0141~\eV$,
$\gamma_{3}/a^{2}=0.0670~\eV$,
$\gamma_{4}/a^{2}=0.232~\eV$,
$\gamma_{5}/a^{2}=0.0070~\eV$.

\subsection{SOC AFM with N\'{e}el vector along $[11\bar{0}0]$}

\begin{table*}
\begin{centering}
\begin{tabular}{|c|c|c|c|c|c|c|c|}
\hline 
 & $E$ & $\{2_{0001}|1/2\}$ & $\{2_{11\bar{2}0}|0\}\mathcal{T}$ & $\mathcal{I}$ &  & $\Gamma$ & $A$\tabularnewline
\hline 
\hline 
$A_{g}$ & 1 & 1 & 1 & 1 & $x^{2}$, $y^{2}$, $z^{2}$, const. & $\sigma_{0}$, $\sigma_{x}$, $\sigma_{0}s_{z}$, $\sigma_{x}s_{z}$ & $\tau_{0,y,z}$, $\sigma_{x}\tau_{0,z}s_{y}$, $\sigma_{x}\tau_{y}s_{y}$\tabularnewline
\hline 
$A_{u}$ & 1 & 1 & 1 & -1 & $z$ &  & $\sigma_{y}\tau_{0,y,z}s_{z}$, $\sigma_{z}\tau_{0,y,z}s_{x}$\tabularnewline
\hline 
$B_{1g}$ & 1 & 1 & -1 & 1 & $xy$ &  & $\tau_{x}$, $\sigma_{x}\tau_{x}s_{y}$\tabularnewline
\hline 
$B_{1u}$ & 1 & 1 & -1 & -1 &  &  & $\sigma_{y}\tau_{x}s_{z}$, $\sigma_{z}\tau_{x}s_{x}$\tabularnewline
\hline 
$B_{3g}$ & 1 & -1 & 1 & 1 & $yz$ & $\sigma_{0}s_{y}$, $\sigma_{x}s_{y}$ & $\sigma_{0}\tau_{0,y,z}s_{y}$, $\sigma_{x}\tau_{0,y,z}s_{0}$\tabularnewline
\hline 
$B_{3u}$ & 1 & -1 & 1 & -1 & $y$ &  & $\sigma_{y}\tau_{x}s_{x}$, $\sigma_{z}\tau_{x}s_{z}$\tabularnewline
\hline 
$B_{2g}$ & 1 & -1 & -1 & 1 & $xz$ & $\sigma_{0}s_{x}$, $\sigma_{x}s_{x}$ & $\sigma_{0}\tau_{x}s_{y}$, $\sigma_{x}\tau_{x}s_{0}$\tabularnewline
\hline 
$B_{2u}$ & 1 & -1 & -1 & -1 & $x$ &  & $\sigma_{y}\tau_{0,y,z}s_{x}$, $\sigma_{z}\tau_{0,y,z}s_{z}$\tabularnewline
\hline 
\end{tabular}
\par\end{centering}
\caption{Irreducible representations of $m'm'm$ magnetic group}
\label{Table:representation_120soc}
\end{table*}

\begin{figure}
\includegraphics[width=1\columnwidth]{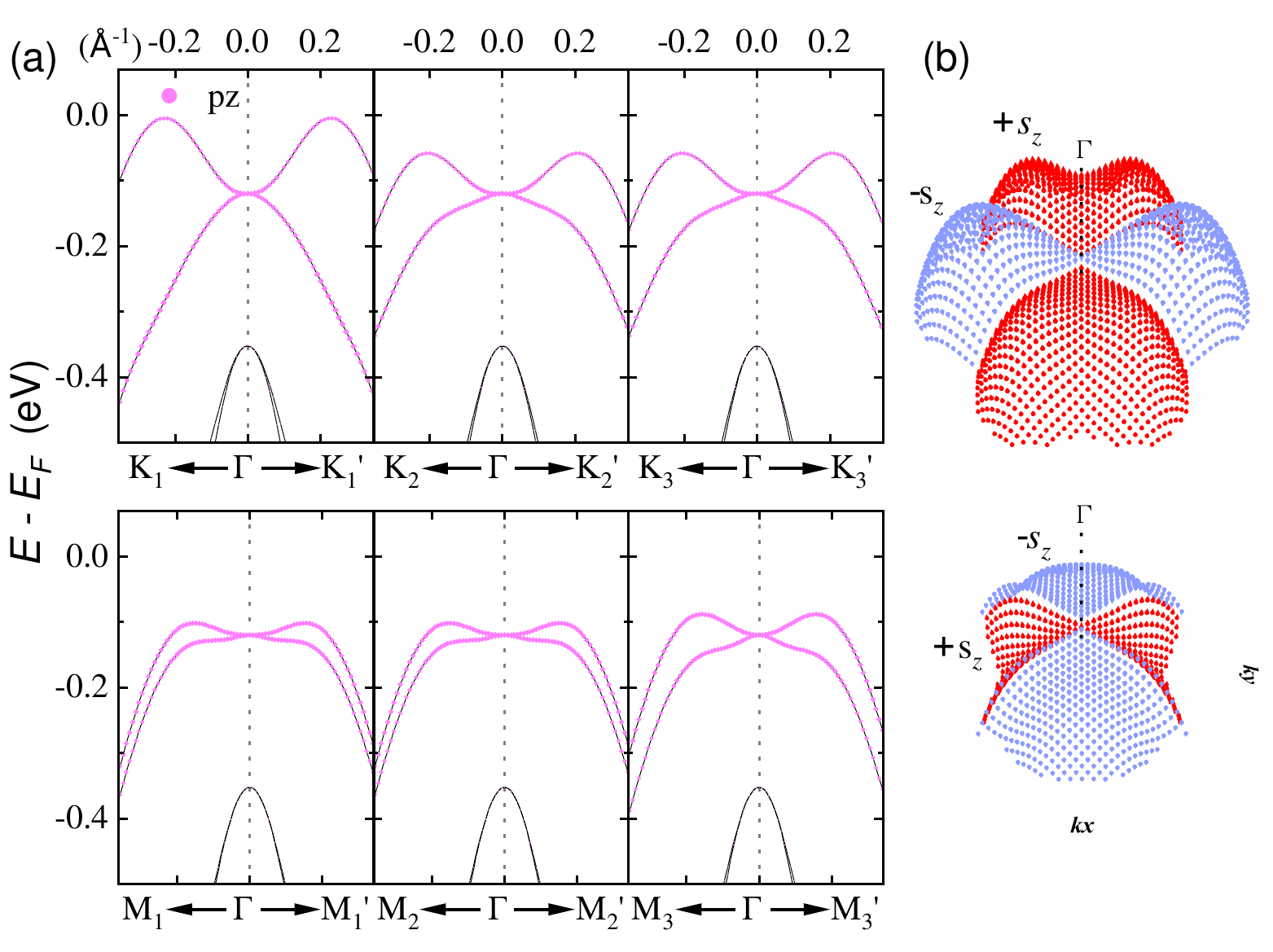}
\includegraphics[width=1\columnwidth]{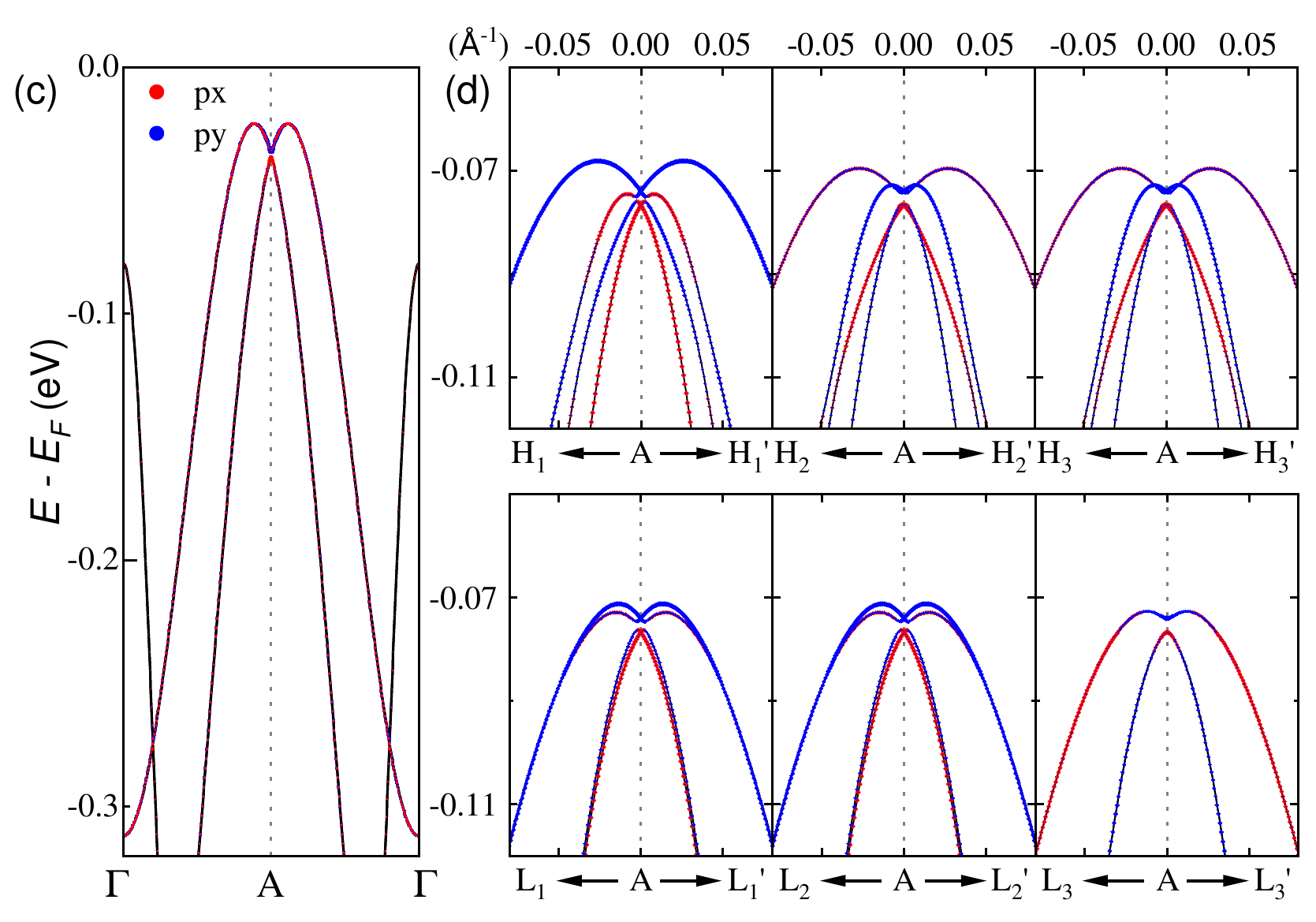}
\caption{With the N\'{e}el vector aligned along the $[1\bar{1}00]$ ($y$) direction, The band structure with Te($5p$) orbital component (a) near the $\Gamma$ point, 
(c) along the $\Gamma$-A-$\Gamma$ $k$-point path, and (d) near A point.
(b) the energy surface for the top two valence bands in $kx$-$ky$ plane near the $\Gamma$ point. 
The red and blue arrows on each dots represent the $+s_z$ and $-s_z$ spin directions respectively.}
\label{fig:af1y} 
\end{figure}

Once the N\'eel vector is along $[1\bar{1}00]$ direction (or $y$ direction),
the SOC terms would be different. 
The fundamental difference compared to $[11\bar{2}0]$ case is the symmetry group. 
In this case, the quotient group $G/T=M$ is a magnetic point group. 
$M=C_{2h}\oplus\mathcal{T}(D_{2h}-C_{2h})=m'm'm$.
It contains $\{1|0\}$, $\mathcal{I}$, $\{2_{0001}|1/2\}$, $\{m_{0001}|1/2\}$, $\{2_{11\bar{2}0}|0\}\mathcal{T}$, $\{2_{1\bar{1}00}|1/2\}\mathcal{T}$,
$\{m_{11\bar{2}0}|0\}\mathcal{T}$, and $\{m_{1\bar{1}00}|1/2\}\mathcal{T}$. 
Generators of the groups are $\{2_{11\bar{2}0}|0\}\mathcal{T}$, $\{2_{0001}|1/2\}$
and $\mathcal{I}$. 

In this case, at the $\Gamma$ point, $\{2_{0001}|1/2\}=\sigma_{x}s_{z}$,
$\{2_{11\bar{2}0}|0\}\mathcal{T}=\sigma_{x}s_{x}is_{y}K=-\sigma_{x}s_{z}K$,
$\mathcal{I}=-\sigma_{x}$. ; at the $A$ point, $\{2_{0001}|1/2\}=i\sigma_{y}\tau_{0}s_{z}$,
$\{2_{11\bar{2}0}|0\}\mathcal{T}=\sigma_{x}\tau_{z}s_{x}is_{y}K=-\sigma_{x}\tau_{z}s_{z}K$,
and $\mathcal{I}=-\sigma_{x}$. The character table is shown in Table \ref{Table:representation_120soc}. 
Remarkably, the trivial irreducible representation at the $\Gamma$ point contains SOC corrections of $\sigma_0s_z$ and $\sigma_xs_z$. 
The effective Hamiltonian in the Hilbert space expanded by the anti-bonding state thus contains terms such as $k_x^2s_z$, $k_y^2s_z$, $k_z^2s_z$. 
Other allowed SOC terms include $k_yk_zs_y$ and $k_xk_zs_x$. 
The tight-binding model suggest that only $k_x^2s_z$ and $k_y^2s_z$ are relevant. 

\begin{align}
H_{[1\bar{1}00]}^{\Gamma} &= H_{\mathrm{non}\mathchar`- \mathrm{SOC}}^{\Gamma} + \gamma_{1}' \left(k_{x}^2 - \lambda_{1}k_{y}^2\right) s_{z} \nonumber \\
&+ \left[\delta_{1}' k_{x}^2 \left(k^2_{x} - \lambda_{2} k^2_{y} \right) + \delta_{2}' k_{y}^2 \left(k^{2}_{x} - \lambda_{2} k^{2}_{y}\right) \right] s_{z} 
\label{eq:gamma_120}
\end{align}
Spins around are thus again polarized along $\pm\hat{z}$ direction. 
Tight-binding results suggest that $\gamma_{1}=2\gamma_{1}'$ and $\lambda_{1}=1$, indicating a $\pi/4$ rotation of the band structure relating to $H_{[11\bar{2}0]}^{\Gamma}$.
The energy surfaces shown in Fig.~\ref{fig:af1y}(b) generally match this rotational relationship.
However, the $[1\bar{1}00]$ case follows different group symmetry with the $[11\bar{2}0]$ case at the $\Gamma$ point, so that there is no constrain of the values of $\gamma_{1},\gamma_{1}'$ and $\lambda_{1}$.
Thus, in detail, the fitting parameters based on DFT band results were obtained as 
$c_{1}/a^2=0.0178~\eV$, 
$c_{2}/c^2=-0.513~\eV$, 
$c_{3}/a^4=-0.0480~\eV$, 
$c_{4}/a^6=0.0045~\eV$, 
$\gamma_{1}'/a^{2}=0.153~\eV$, 
$\delta_{1}'/a^{4}=0.0571~\eV$, 
$\delta_{2}'/a^{4}=0.0273~\eV$,
$\lambda_{1}=1.31$ and $\lambda_{2}=6.15$.
$\lambda_{1}$ is not 1 and $\lambda_{2}$ is not equal to $\lambda_{1}$.
This indicate that the degenerate lines with zero for all the $s_{z}$ term are not the simple straight lines $k_{x}=\pm k_{y}$ but arcs with mirror symmetries along $k_{x}=0$ and $k_{y}=0$.

At the $A$ point, $\ket{+}$ and $\ket{-}$ states, the spin part for the two basis for $\mathbf{\omega}$ matrices are now determined by the eigenstates of $s_y$ instead, i.e., $s_y\ket{\pm}=\pm\ket{\pm}$.
In contrast to the case at the $\Gamma$ point, many terms are allowed at the $A$ point, that include 
$\tau_{y,z}$, 
$k_{x}k_{y}\tau_{x}$, 
$k_{y}k_{z}\tau_{0,y,z}\omega_{z}$, 
$k_{x}k_{z}\tau_{x}\omega_{z}$, 
$k_{z}\tau_{0,y,z}\omega_{y}$, 
$k_{y}\tau_{x}\omega_{x}$, 
and $k_{x}\tau_{0,y,z}\omega_{x}$. 
The tight-binding suggests the following effective Hamiltonian
\begin{eqnarray}
    H_{[1\bar{1}00]}^{A} &= H_{\mathrm{non}\mathchar`- \mathrm{SOC}}^{A} - \gamma_{0}' \tau_{z}  + \gamma_{1}' \left(k_{x} \tau_{z} -  k_{y}  \tau_{x}\right) \omega_{x} - \gamma_{2}' k_{x} \tau_{y}  \omega_{x}  \nonumber \\ 
    &+ \gamma_{3}' k_{z} \tau_{y}  \omega_{y} + \gamma_{4}'/2 \left(k_{x}^{2} - k_{y}^{2}\right) \tau_{y} + \gamma_{5}' k_{y} k_{z} \tau_{y} \omega_{z}
\label{eq:A_120}
\end{eqnarray}
The fitting parameters from the DFT bands are
$c_{1}/a^2=1.089~\eV$, 
$c_{2}/c^2=-0.132~\eV$, 
$c_{3}/a^2=-0.615~\eV$, 
$c_{4}/a^2=-0.0011~\eV$, 
$\gamma_{0}'=0.0013~\eV$, 
$\gamma_{1}'/a^{2}=-0.103~\eV$, 
$\gamma_{2}'/a^{2}=-0.0591~\eV$,
$\gamma_{3}'/a^{2}=0.0702~\eV$,
$\gamma_{4}'/a^{2}=0.299~\eV$,
$\gamma_{5}'/a^{2}=0.0054~\eV$.
The zero-ordered term  $\gamma_{0}=\gamma_{0}'$ is confirmed by both tight-binding model and the DFT bands.
$\gamma_{4}\approx\gamma_{4}'$ also indicates that $H_{[1\bar{1}00]}^{A}$ generally follows a $\pi/4$ rotation of the SOC part of the band structure relating to $H_{[11\bar{2}0]}^{A}$.
The first-ordered coefficients $\gamma_{2}/\gamma_{2}'$ and $\gamma_{3}/\gamma_{3}'$ and the second-ordered one $\gamma_{5}/\gamma_{5}'$ determines the anisotropic dispersion away from non-SOC $C_3$ symmetry.
In both cases, the magnitudes of these coefficients are at least one-order of magnitude smaller than those of the corresponding ordered coefficients at the $\Gamma$ point.
Both the DFT bands [Fig.~\ref{fig:af1x}(d) and Fig.~\ref{fig:af1y}(d)] and the band fitting results display the tiny anisotropic behavior within several $\meV$.

\section{Fermi pockets at the $\Gamma$ and $A$ point}

\begin{figure}
\includegraphics[width=1\columnwidth]{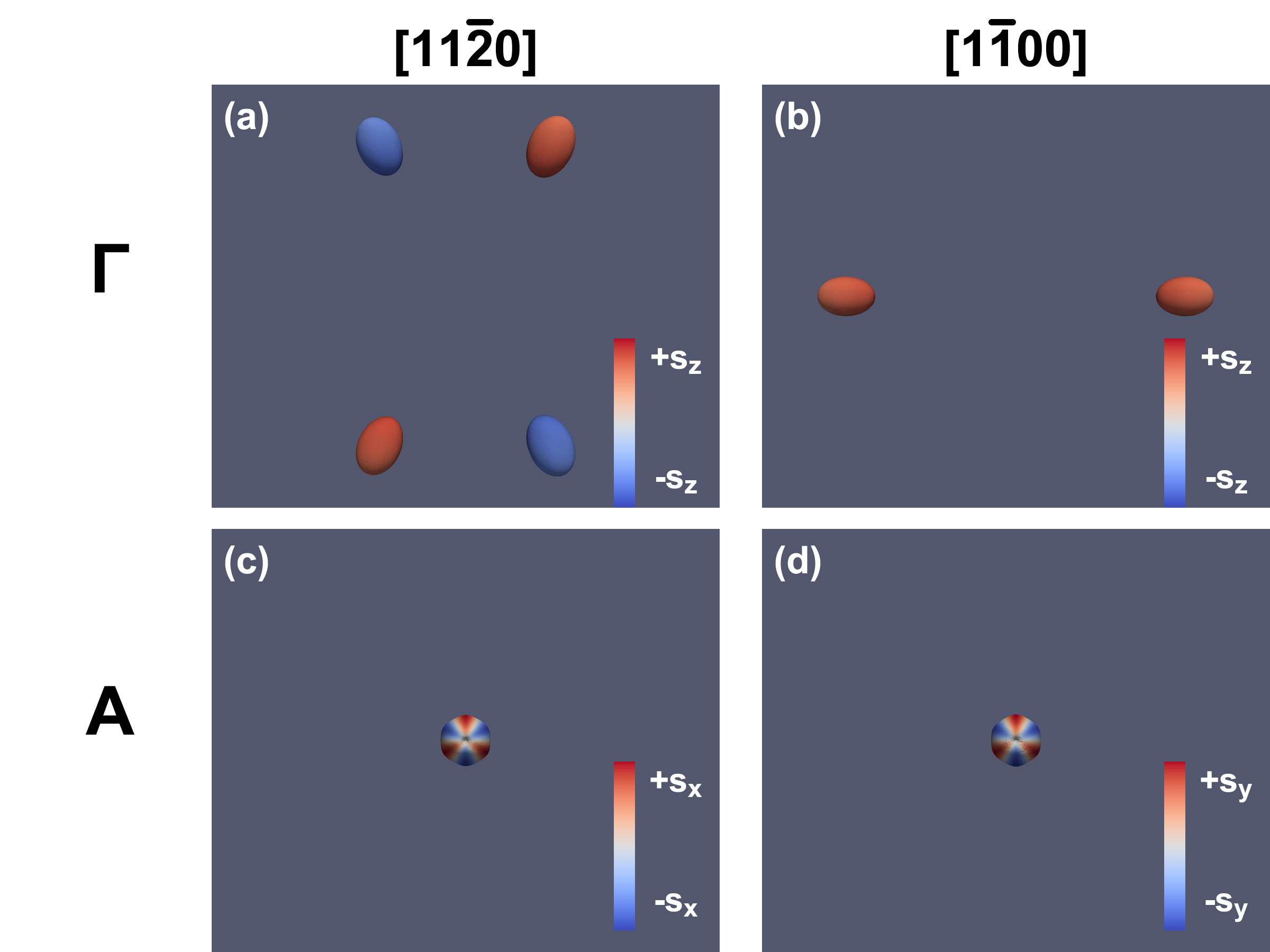}
\caption{Fermi surfaces in four cases: around the $\Gamma$ point, N\'{e}el vector along (a) $[11\bar{2}0]$ and (b) $[1\bar{1}00]$ respectively; around the $A$ point, N\'{e}el vector along (c) $[11\bar{2}0]$ and (d) $[1\bar{1}00]$ respectively. The Fermi energies for all the four cases are $0.01~\eV$ lower than the VBM. Red and blue color represents the spin components.}
\label{fig:soc-fs}
\end{figure}

It is very difficult to determine whether whether the VBM of MnTe is located around the $\Gamma$ or the $A$ point
since their relative energy difference is highly sensitive to external environments including strain, impurities, temperature, etc. 
The calculated position of VBM is also sensitive to the exchange-correlated functionals\cite{YinGen_Theory_MnTe}. 
Here, we focus on the shape and spin properties of the Fermi surfaces when the VBM is located around the $\Gamma$ and the $A$ point respectively under $p$-doping.

Once the VBM is located around the $\Gamma$, the Fermi surfaces look significantly different when the N\'{e}el vector is aligned along the $[11\bar{2}0]$ and $[1\bar{1}00]$ direction, as shown in Fig.~\ref{fig:soc-fs}(a)(b).
In the case of $[11\bar{2}0]$, four VBM $k$-points are presented according to the energy surfaces shown in Fig.~\ref{fig:af1x}(b). 
Under $p$-doping environment, four hole pockets are symmetrically distributed across the four quadrants, mirroring each other with respect to $k_x=0$ and $k_y=0$ [Fig.~\ref{fig:soc-fs}(a)].
The pockets in I/III quadrants has the opposite spin direction to those in II/IV quadrants and all four pockets are spin polarized along $s_z$ direction.
The net magnetization of the Fermi surfaces is zero,
but the anisotropic distribution of the Fermi surfaces leads to the anisotropic conductivity.
The authors' earlier work~\cite{YinGen_Theory_MnTe} predicted giant $\mathcal{T}$-even planer Hall effect with the maximum Hall angle near $30\%$ for this case of Fermi surfaces.

In the case of $[1\bar{1}00]$, the energy maximum occur at two points, which are located on the line $k_{y}=0$ and are symmetric with respect to the $\Gamma$ point, as shown in Fig.~\ref{fig:af1y}(b).
Thus, only two corresponding hole pockets appears symmetrically at $+x$ and $-x$ regions along $k_{y}=0$ [Fig.~\ref{fig:soc-fs}(b)].
Surprisingly, both pockets have the same spin polarization along $s_z$ direction so that the valence electron is fully spin polarized even though it is a collinear antiferromagnet. 
This is consistent with the effective Hamiltonian in Eq.~\ref{eq:gamma_120}. 
All spin-dependent terms therein are $s_z$ related, and the coefficients are even functions of both $k_x$ and $k_y$.
The anisotropic and spin polarized Fermi surfaces is originate from SOC,
and bring about not only the anisotropic conductivity but also the spin-polarized current.
Strong spin-transfer torque with strong spin scattering between itinerant electrons polarized along $s_z$ given local magnetic moments on Mn along $s_y$ are expected.
The strong planer Hall response is also expected in terms of the anisotropic Fermi surfaces.
In both cases, the energy window for the strong anisotropic Fermi surfaces is about $0.08~\eV$ which is adequate for the $p$-doping to against thermal fluctuation.

When the VBM is located around $A$, the valence bands for the two cases of N\'{e}el vector appear very similar since the impact from SOC is limited.
To this end, in both cases, the Fermi surface generally exhibits a hexagram structure, as shown in Fig.~\ref{fig:soc-fs}.
The spin component, following the non-SOC band structure, shows the bulk $g$-wave spin splitting at low symmetry areas, 
and the spin direction follows the N\'{e}el vector direction,
so that the spin properties are dominated by the non-relativistic altermagnetic feature.
The energy window is no more than $0.01~\eV$ to establish the altermagnetic spin splitting feature such as intrinsic anomalous Hall effect in spintronic measurement, 
so that it is highly susceptible to external environmental factors such as thermal fluctuation and impurities.
It is consistent with the experimental measurement of the spontaneous anomalous Hall conductivity of MnTe as only $\sim0.02~\mathrm{S/cm}$~\cite{Gonzalez_2023}. 

\section{Conclusion}

In summary, we resolved the band structures of $\alpha$-MnTe near the VBM at the $\Gamma$ point and the $A$ point.
The group representation theory, first-principles calculations, the tight-binding theory give consistent effective Hamiltonian around high symmetry points of MnTe. Compared to direct $k\cdot p$ theory, our model Hamiltonian is built upon realistic bases. We showed that spin-orbit coupling is essential in generating spin splitting at high symmetric points and high symmetric lines. The effective Hamiltonian could be used in future studies of electron, spin, and orbital transports of MnTe and other transition metal chalcogenides with the same structure. The spin polarized Fermi pockets could give nontrivial magnetoresistance and even possible electron pairing states. 

\bibliography{main}	

\end{document}